\documentclass[showpacs,amsmath,amssymb,aps,prb,10pt,reprint,superscriptaddress]{revtex4-2}
\usepackage{bm}
\usepackage[breaklinks=true,colorlinks=true,linkcolor=blue,urlcolor=blue,citecolor=blue]{hyperref}
\usepackage{dcolumn}
\usepackage{amsmath,amssymb}
\usepackage{mathdots}
\usepackage{natbib}
\usepackage{soul,color}
\usepackage{graphicx}
\usepackage[note-name]{notes2bib}
\usepackage{mathptmx}
\usepackage[margin=1in]{geometry}

\begin{document}

\title{Crystal structure effects on vortex dynamics in superconducting MgB$_2$ thin films}

\author{Clemens Schmid}
    \email[Corresponding author: ]{c.schmid@univie.ac.at}
    \affiliation{Faculty of Physics, University of Vienna, Austria}
    \affiliation{Vienna Doctoral School in Physics, Vienna, Austria}

\author{Anton Pokusinskyi}
    \affiliation{Cryogenic Quantum Electronics, Institute for Electrical Measurement Science and Fundamental Electrical Engineering (EMG) and Laboratory for Emerging Nanometrology (LENA), Technische Universit\"at Braunschweig, 38106, Braunschweig, Germany}

\author{Markus Gruber}
    \affiliation{Faculty of Physics, University of Vienna, Austria}
    
\author{Corentin Pfaff}
    \affiliation{Institute Jean Lamour, Université de Lorraine-CNRS, Nancy, France}

\author{Theo Courtois}
    \affiliation{Institute Jean Lamour, Université de Lorraine-CNRS, Nancy, France}

\author{Alexander Kasatkin}
    \affiliation{G.V. Kurdyumov Institute for Metal Physics of the NAS of Ukraine, 03142, Kyiv, Ukraine}

\author{Karine Dumesnil}
    \affiliation{Institute Jean Lamour, Université de Lorraine-CNRS, Nancy, France}

\author{Stephane Mangin}
    \affiliation{Institute Jean Lamour, Université de Lorraine-CNRS, Nancy, France}

\author{Thomas Hauet}
    \affiliation{Institute Jean Lamour, Université de Lorraine-CNRS, Nancy, France}

\author{Oleksandr Dobrovolskiy}
    \affiliation{Cryogenic Quantum Electronics, Institute for Electrical Measurement Science and Fundamental Electrical Engineering (EMG) and Laboratory for Emerging Nanometrology (LENA), Technische Universit\"at Braunschweig, 38106, Braunschweig, Germany}
    \affiliation{FLUXONICS---The European Foundry for Superconducting Electronics e.V., 38116 Braunschweig, Germany}

\date{\today}

\begin{abstract}
The current-driven resistive transition is central to superconducting single-photon detectors, transition-edge sensors, and fluxonic devices. Depending on sample uniformity, dimensions, and heat removal, it can be driven by phase-slip events, flux-flow instabilities (FFI), or normal-domain formation. Here, we investigate the influence of two types of microstructural defects on vortex dynamics in MgB$_2$ films: columnar growth in textured films and buffer-layer roughness in single-crystal films. The current-voltage ($I$-$V$) curves measured at $T \approx 0.25 T_\mathrm{c}$ for both films exhibit multiple steps. Time-dependent Ginzburg-Landau simulations reproduce the major features of the experimental $I$-$V$ curves and suggest that the resistive transitions for both films are mediated by the formation and growth of normal domains rather than FFI. The single-crystal film with buffer-layer roughness exhibits superconductivity breakdown at higher currents and pinning activation energies approximately twice those of the textured film, along with more pronounced multi-step features in the $I$-$V$ curves. These features are attributed to the combination of stronger pinning induced by lateral variations of the superconducting order parameter along the MgO buffer layer and its lower thermal boundary resistance. Our results show that both the film microstructure and the film-buffer interface are critical for the resistive transition, offering insights for superconducting devices requiring controlled dissipation at high transport currents.
\end{abstract}
\maketitle

\maketitle

\section{Introduction}
The current-induced transition to the high-resistive or normal state is central to superconducting single-photon detectors (SSPDs) \cite{Gol01apl}, transition-edge sensors \cite{Loe19acs}, and various Josephson \cite{Cat25nac} and Abrikosov \cite{Dob20nac} fluxonic devices. Within the Ginzburg-Landau framework, the maximal dissipationless current in superconductors is set by the depairing current density $j_\mathrm{dep}$\,\cite{Tin04boo}. In practice, however, superconductivity is typically destroyed at a critical current density $j_\mathrm{c}$ \cite{Rui26pms} well below this theoretical limit\,\cite{Vod17pra}. The primary mechanisms driving the transition to a high-resistive state include the increase of the phase difference across weak links under overcritical currents \cite{Bar82boo}, formation and growth of normal (N) domains \cite{Rey02prb}, flux-flow instabilities (FFI)  \cite{Dob24cmp}, slips of the phase of the superconducting order parameter \cite{Siv03prl}, and their combinations \cite{Ada15prb,Emb17nac,Bez19prb,Bev23pra}. Which of the mechanisms dominates depends on the sample uniformity and dimensions relative to the coherence length $\xi$ and magnetic penetration depth $\lambda$ \cite{Ivl84phb}, and the rates of electron energy relaxation and heat removal \cite{Sid18prb}. In the case of FFIs, the current-voltage ($I$-$V$) curve of a superconductor exhibits an abrupt jump \cite{Lar76etp2} while for N domains \cite{Rey02prb} and phase slips \cite{Buh15nac} multiple voltage transitions are usually observed \cite{Siv03prl,Zol14ltp}.

In the FFI framework, the voltage $V^\ast$ measured just before the resistance jump defines the maximal vortex velocity, $v^\ast = V^\ast/(B L)$, where $B$ is the magnetic flux density and $L$ the voltage-lead separation. Near $T_\mathrm{c}$, FFI is electronic~\cite{Lar76etp2}, whereas at lower temperatures, electron-electron scattering dominates and FFI becomes thermal~\cite{Kunchur2002}. A larger $v^\ast$ implies a shorter energy-relaxation time $\tau_\varepsilon$, a desirable property for SSPDs\,\cite{Vod17pra}. In uniform systems, the highest $v^\ast$ of a few tens of km/s were experimentally observed in Nb-C\,\cite{Dob20nac} and MoSi\,\cite{Bud22pra} strips with small electron diffusion coefficient $D\simeq0.5$\,cm/s$^2$ and fast electron energy relaxation ($\tau_\varepsilon\simeq 20$-$30$\,ps) and in Pb bridges \cite{Emb17nac} with large $D\simeq30$\,cm/s$^2$ and short inelastic electron-phonon relaxation time ($\tau_\mathrm{ep}\simeq 20$\,ps)\,\cite{Wat81ltp}.

However, in many superconductors relevant for SSPDs, the $v^\ast$ and $\tau_\varepsilon$ values inferred from FFI often conflict with the fast relaxation observed in photon-counting experiments~\cite{Bud22pra}. This discrepancy arises because FFI may nucleate locally rather than across the entire sample~\cite{Bez19prb}. In inhomogeneous systems, regions with slow and fast vortices can coexist~\cite{Sil12njp,Ada15prb}, leading to the formation of N domains where $v^\ast$ is first reached and rendering the extracted $v^\ast$ non-quantitative. 

Once the current density exceeds the threshold\,\cite{Gur84spu,Bez84ltp} 
\begin{equation}
    \label{eq:jeq}
    j_\mathrm{eq} = \biggl[\frac{2 h}{R_\square d^2} (T_\mathrm{c} - T)\biggr]^{1/2},
\end{equation}
which determines the equilibrium of the nonisothermal normal/superconductor (N/S) boundary, domains will expand until the entire sample transitions to the normal state \cite{Bez19prb}. In Eq.\,\eqref{eq:jeq}, $h$ is the heat-removal coefficient, $R_\square$ the film sheet resistance in the normal state, and $d$ the film thickness. By contrast, if the transport current density $j < j_\mathrm{eq}$, a non-stationary N domain can appear in the film: it initially grows, reaches a maximal size, and then shrinks until it vanishes \cite{Bez84ltp}. In the presence of N domains, the relation $v^\ast = V^\ast/(B L)$ no longer provides a quantitative measure of the instability velocity. At the same time, a crossover from local to global FFI can be facilitated by vortex ordering through a reduction of intrinsic random disorder \cite{Sil12njp}, vortex guiding effects \cite{Dob19pra}, and coupling to spin waves \cite{Dob25nan}. High-quality sample edges are also crucial \cite{Sil25apr}, as they promote FFI nucleation at multiple points along the sample border \cite{Bud22pra}, thereby mitigating local overheating \cite{Bez19prb}. 

Here, we investigate the impact of two defect types---interfacial and volumetric crystal defects---on vortex pinning, dynamics, and resistive states in single-crystal MgB$_2$ films with rough MgO buffer layers and in textured MgB$_2$ films grown on atomically flat amorphous MgBAlO$_\mathrm{x}$ buffers. MgB$_2$ is a two-band superconductor~\cite{Nag01nat,Fer01sst,Zha13tas}, with $T_\mathrm{c}$ exceeding 20\,K in films just a few tens of nanometers thick and reaching 39\,K in bulk samples. Recently, MgB$_2$ films have enabled SSPDs up to $20$\,K~\cite{Cha24nac}, high-kinetic-inductance devices~\cite{Gre25apl}, and spin injection at MgB$_2$/ferromagnet interfaces~\cite{Pfa24apl}. Yet, vortex dynamics and resistive states in MgB$_2$ films at high transport currents remain largely unexplored. Here, we bridge this gap by measuring $I$-$V$ curves of single-crystal and textured MgB$_2$ films at $T \approx 0.25\,T_\mathrm{c}$, both exhibiting multi-step characteristics at low magnetic fields. Based on a comparison of our experimental data with time-dependent Ginzburg-Landau simulations, we rule out an FFI scenario and attribute the resistive behavior to the nucleation and growth of N domains. We then examine how the two types of structural defects, along with heat removal through the buffer layer, influence pinning activation energy, transition currents, and N-domain evolution.

\section{Samples} 
MgB$_2$ thin films were grown by molecular beam epitaxy under ultra-high-vacuum conditions in a chamber with a base pressure in the $10^{-10}$ Torr range \cite{Pfa24apl}. The films were grown on c-cut sapphire substrates. Prior to film growth, the substrates were annealed at $1000\,^\circ$C to obtain clean, well-ordered surfaces. During deposition, the chamber pressure was maintained at approximately $10^{-8}$ Torr. Magnesium was evaporated from an effusion cell, while boron was evaporated using an electron-beam gun. Evaporation rates were controlled using a quartz crystal microbalance. Reflection high-energy electron diffraction (RHEED) was performed \textit{in situ} to monitor the crystalline quality of the films. The deposition temperature $T_\textrm{s}$ was monitored using either a pyrometer or a thermocouple attached to the sample holder. After systematically varying $T_\mathrm{s}$ and the deposition rates, optimal growth was achieved at $T_\mathrm{s} \approx 370\,^\circ\mathrm{C}$ with Mg and B rates of $1$ and $0.1$\,\AA/s, respectively, yielding an effective Mg:B ratio of $3{:}1$ to compensate for the higher re-evaporation rate of Mg \cite{Ueda2003,Xi2009,Li2017}. Lowering the deposition temperature by a few tens of degrees forms Mg clusters from insufficient desorption, while its increasing induces multiphase polycrystalline growth. \textit{In situ} x-ray photoemission spectroscopy confirmed that the MgB$_2$ surface is free of oxygen and carbon.

\begin{figure}
     \centering
     \includegraphics[width=\linewidth]{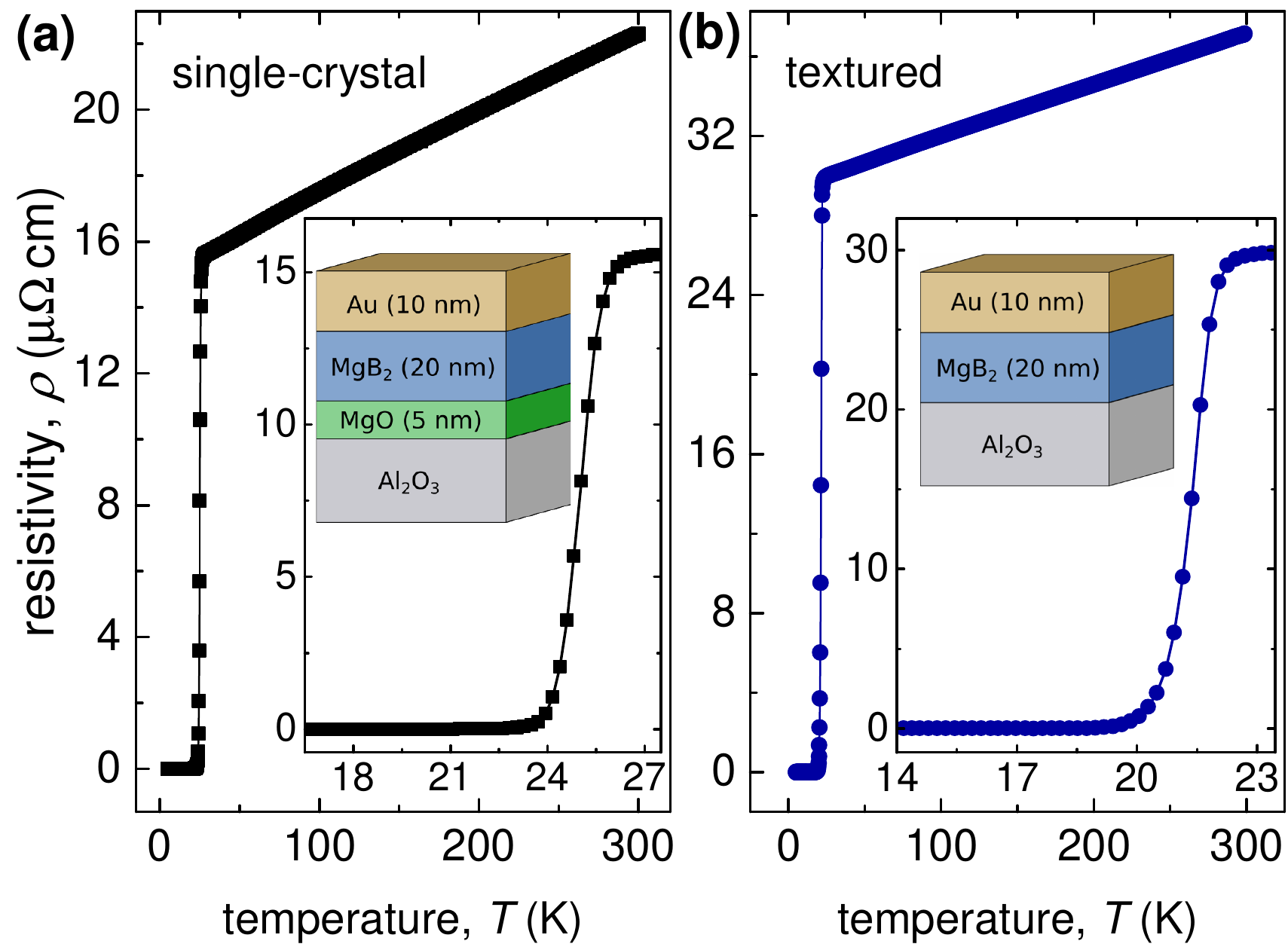}
     \\[1mm]
     \includegraphics[width=\linewidth]{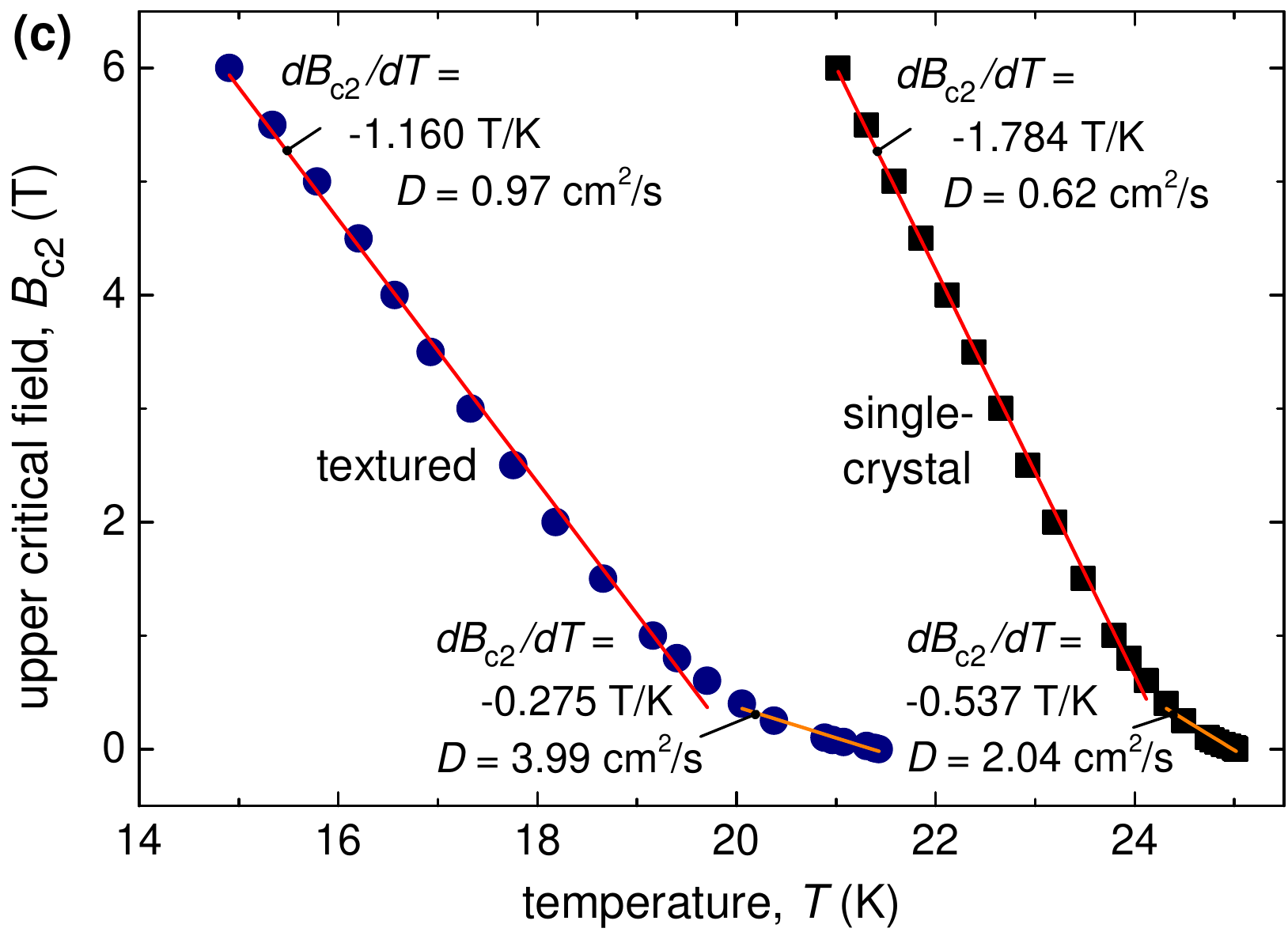}
        \caption{Temperature dependence of the resistivity for the single-crystal (a) and textured (b) MgB$_2$ films. Insets: stack structures and enlarged view of the superconducting transitions. (c) Temperature dependence of the upper critical field for both samples. Symbols: experimental data; solid lines: linear fits.}
    \label{fig:cooling}
\end{figure}

For this study we used one textured and one single-crystal MgB$_2$ film, each with a thickness $d$ of $20$\,nm, hereafter referred to as sample T and sample S, respectively. The single-crystal film was grown on a 4\,nm-thick MgO (111) buffer layer, deposited at $900\,^\circ$C on the sapphire substrate prior to MgB$_2$ deposition. The chemical stability of MgO and the small lattice mismatch of approximately $3.7\%$ between MgO (111) and MgB$_2$ (0001) enable the growth of epitaxial, single-crystal MgB$_2$ films \cite{Li2017,Pfa24apl}. In contrast, direct growth of MgB$_2$ on the sapphire substrate yielded a textured film. Both films were then capped \textit{in-situ} with an 8\,nm-thick gold layer to prevent oxidation. The layer structure of the fabricated stacks is illustrated in the insets of Fig. \ref{fig:cooling}(a,\,b).

\begin{figure}
    \includegraphics[width=1\linewidth]{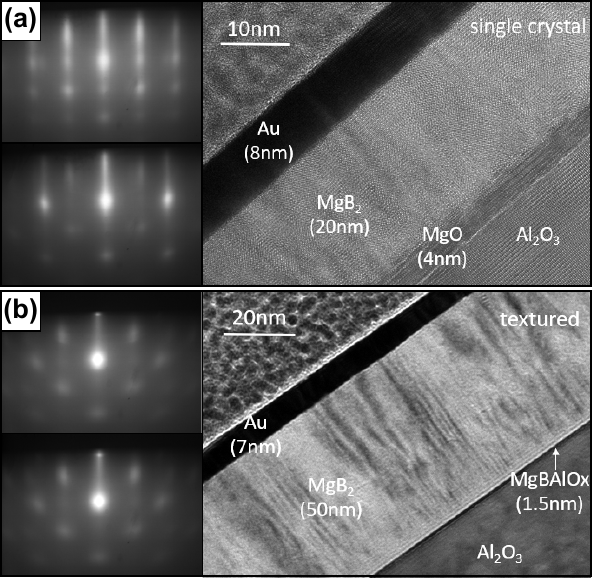}
    \caption{RHEED patterns obtained upon sample in-plane rotation by $30^\circ$ (left panels) and HRTEM images (right panels) for the single-crystal (a) and textured (b) MgB$_2$ films.}
    \label{fig:structure}
\end{figure}

Figure \ref{fig:structure} shows RHEED diffraction patterns and high-resolution transmission electron microscopy (HRTEM) images of  a single-crystal (a) and a textured (b) MgB$_2$ film. Note that Fig.~\ref{fig:structure}(b) presents an HRTEM image of a 50\,nm-thick textured sample, rather than the 20\,nm-thick sample T, in order to better highlight the columnar growth and its expansion through the film thickness. The ordered arrays of diffraction spots and streaks in the RHEED patterns indicate well-ordered growth for both films. For sample S, in-plane rotation by $30^\circ$ in Fig.\,\ref{fig:structure}(a) reveals a difference in the RHEED pattern, pointing to a fixed epitaxial relationship of the film with the substrate. In contrast, the invariance of the spots upon in-plane rotation for sample T in Fig.\,\ref{fig:structure}(b) indicates a textured film with a common c-axis orientation of crystallites perpendicular to the film plane.

The HRTEM image of sample S in Fig.\,\ref{fig:structure}(a) confirms ordered film growth with the epitaxial relationship [1-210] Al$_2$O$_3$ (0001)/[2-1-1] MgO (111)/[10-10] MgB$_2$ (0001). The $c$-axis parameter of the MgB$_2$ hexagonal lattice is $3.54$\,\r{A}, close to the bulk value \cite{Jorgensen2001}. Overall, the MgO/MgB$_2$ interface exhibits low contrast, indicating a gradual variation of the film/buffer layer properties. Its rms roughness over a 100\,nm scan is approximately 2\,nm, with typical lateral contrast variations on a length scale of about $6$-$8$\,nm. The MgB$_2$/Al$_2$O$_3$ interface in sample T, shown in Fig.\,\ref{fig:structure}(b), appears much sharper. The textured MgB$_2$ film shows columnar structures with diameters below $1$\,nm near the Al$_2$O$_3$/MgB$_2$ interface and their diameters increase up to about $5$\,nm at the MgB$_2$/Au interface. The film's polycrystallinity is attributed to the large lattice mismatch ($>10$\,\%) between the two materials and to the presence of an intermixed layer at the Al$_2$O$_3$/MgB$_2$ interface \cite{Tian2002, Jo2002, Saito2004, Gu2005}. This atomically smooth amorphous intermediate layer is about $1.5$\,nm thick, and energy-dispersive x-ray (EDX) spectroscopy (not shown) reveals that it consists of a mixture of Al, O, Mg, and B atoms. Above the interfacial layers, EDX confirms the MgB$_2$ stoichiometry for both samples. The rms roughness at the MgB$_2$/Au interface does not exceed $1$\,nm over $100$\,nm for either film.

\section{Resistance measurements}

Both films were patterned into standard four-probe geometries using conventional UV lithography for electrical resistance measurements. The resulting microstrips have a width $W$ of $10\,\mu$m and a voltage-lead separation $L$ of $28\,\mu$m. The full layer stacks were first patterned and then etched down to the substrate by ion beam milling, with the milled thickness monitored using an ion mass spectrometer. In a second step, Ti(10\,nm)/Au(150\,nm) contacts were patterned and deposited using an e-beam evaporator. Resistance measurements were performed in a Quantum Design PPMS with magnetic field perpendicular to the film plane. $I$-$V$ curves were acquired in the current-driven regime with up-sweeps of the dc current.

\subsection{Superconducting parameters}

The temperature dependence of the resistivity, $\rho(T)$, for both films is shown in Fig.\,\ref{fig:cooling}. At $30$\,K, sample S has a resistivity of $\rho_\mathrm{30\,K} = 16\,\mu\Omega$cm, while for sample T it is $30\,\mu\Omega$cm. The residual resistivity ratios (RRR), defined as the ratio of the resistivity at $300$\,K to that at $30$\,K, are $1.4$ and $1.25$ for samples S and T, respectively. The insets in Fig.\,\ref{fig:cooling} display the $\rho(T)$ curves through the superconducting transition, with $T_\mathrm{c} = 25.0$\,K for sample S and $21.6$\,K for sample T. The transition temperatures were determined using the $50\%$ resistance criterion. The transition widths $\Delta T_\mathrm{c}$ are $0.5$\,K for sample S and $1$\,K for sample T. The higher resistivity and lower $T_\mathrm{c}$ are ascribed to the polycrystalline structure of sample T.

Applying a magnetic field broadens the superconducting transition and systematically shifts it to lower temperatures. Near $T_\mathrm{c}$, the temperature dependence of the upper critical field $B_\mathrm{c2}(T)$ exhibits two slopes $dB_{\mathrm{c}2}/dT =-1.78$\,T/K and $dB_{\mathrm{c}2}/dT =-0.54$\,T/K for sample S, see Fig.\,\ref{fig:cooling}(c).
 For sample T, these values are $dB_{\mathrm{c}2}/dT =-1.16$\,T/K and $dB_{\mathrm{c}2}/dT =-0.28$\,T/K.
These slopes correspond to electron diffusion coefficients $D$ of $0.62$\,cm$^2$/s and $2.04$\,cm$^2$/s for sample S, and to $0.97$\,cm$^2$/s and $3.99$\,cm$^2$/s for sample T. The diffusion coefficients were deduced using the relation $D = -1.097(dB_{\mathrm{c}2}/dT)^{-1}|_{T = T_\mathrm{c}}$ \cite{Tin04boo}. The two slopes in $B_\mathrm{c2}(T)$ for MgB$2$ were observed previously \cite{Sza02pcs} and associated with the two-gap nature of the superconducting order parameter \cite{Dah03prl}. Near $T_\mathrm{c}$, both bands contribute, and the more isotropic $\pi$-band with higher diffusivity plays a stronger role \cite{Nag01nat,Fer01sst,Zha13tas}. At lower temperatures, the $\sigma$-band with stronger pairing and lower diffusivity dominates. In the following, measurements were performed well below $T_\mathrm{c}$, in the regime where the $\sigma$ band dominates vortex dynamics and critical currents. The coherence lengths at zero temperature are estimated\,\cite{Bud22pra} as $\xi(0) = \sqrt{\hbar D/1.76k_\mathrm{B}T_\mathrm{c}} = 3.3$\,nm and $6$\,nm for sample S, and $4.4$\,nm and $9$\,nm for sample T. The magnetic penetration depth is estimated as $\lambda(0) = 1.05\cdot10^{-3} \sqrt{\rho_\mathrm{30\,K} /T_\mathrm{c}} \approx 90\,$nm, with the Pearl length $\Lambda(0) = 2\lambda^2(0)/d \approx 800$\,nm for sample S. For sample T, we obtain $\lambda(0) = 125$\,nm and $\Lambda \approx 1.5\,\mu$m. Thus, our strips are thin, with $d \ll \lambda(0)$, and wide, with $\xi \ll  \Lambda(0) \ll W$.

\subsection{Arrhenius analysis}
\label{sec:Arrhenius}

\begin{figure}
     \centering
    \includegraphics[width=\linewidth]{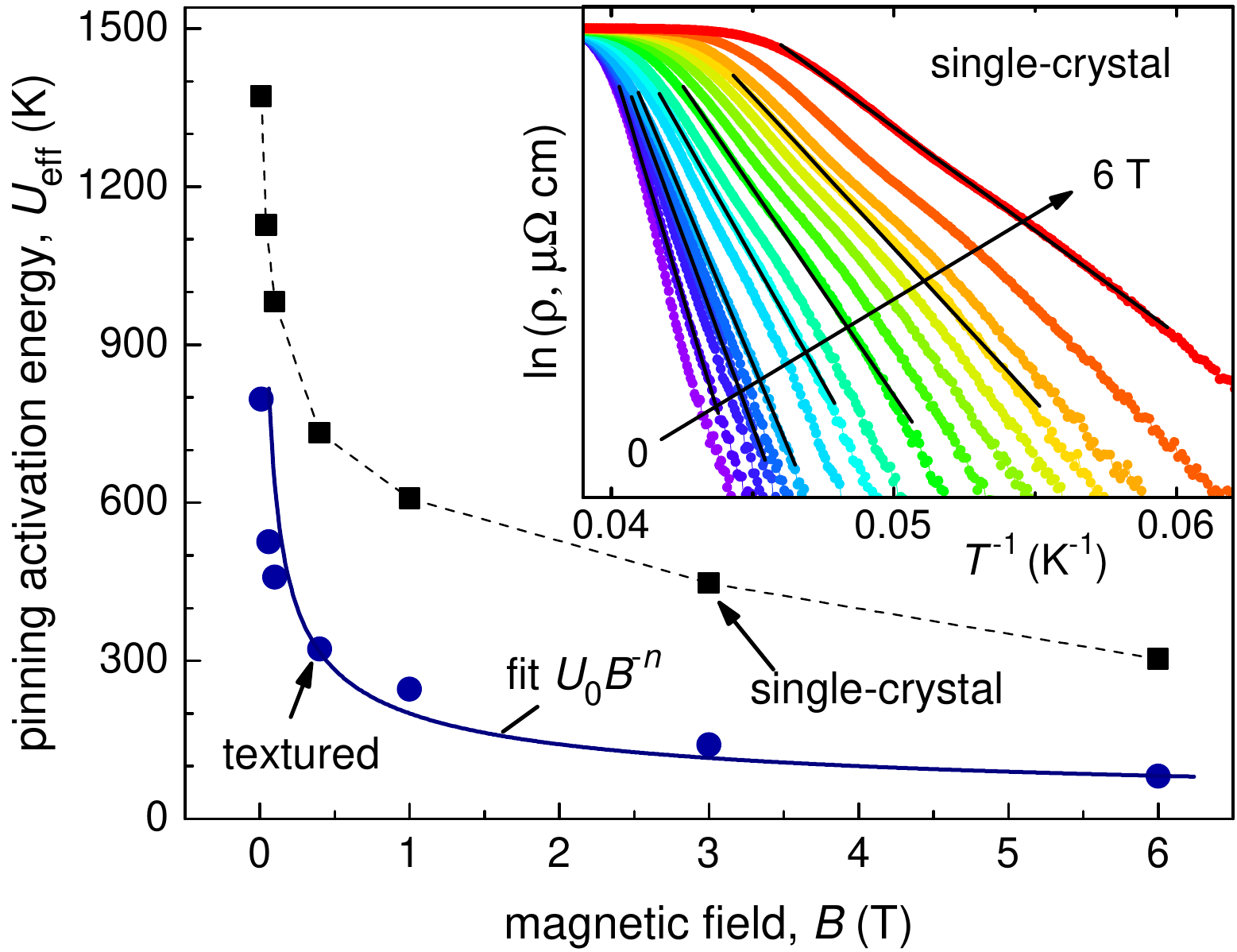}
     \caption{Pinning activation energy for the single-crystal and textured MgB$_2$ films as a function of the applied magnetic field. Symbols: experimental data; dashed line: guide to the eye; solid line: fit to $U_\mathrm{eff} = U_0 B^{-n}$ with $U_0 = 200$\,K and $n = 0.5$. Inset: Representative Arrhenius plot from which the activation energies of the single-crystal film were deduced using linear fits at the corresponding magnetic fields.}
     \label{fig:Arrhenius}
\end{figure}

In order to quantify the pinning potential that prevents vortex motion, an Arrhenius analysis of the resistivity has been performed. At small transport currents, the resistivity follows an Arrhenius law, $\rho = \rho_0 \exp(-U_\mathrm{eff}/k_\mathrm{B} T)$, where $\rho_0$ is a prefactor and $U_\mathrm{eff}$ is the effective pinning activation energy \cite{Tsukui2001,Dobrovolskiy}. The analysis was carried out for magnetic fields up to $6$\,T. The deduced activation energies for both films are presented in Fig.\,\ref{fig:Arrhenius}, with an Arrhenius plot for sample~S shown in the inset. The systematically lower $U_\mathrm{eff}$ values for sample~T---reduced by a factor of two to three compared to sample~S---indicate weaker pinning, as smaller energy barriers suffice to activate vortex motion. At first glance, this observation may appear counterintuitive, since the single-crystal film is expected to have a low density of structural defects. However, structural perfection alone does not guarantee weak overall pinning, which also arises from volume defects, geometrical constrictions, edge quality, thickness variations, and suppression of the superconducting order parameter at the film-substrate and film-capping interfaces \cite{Rui26pms}.

Since the well-ordered crystal structure of sample S implies weak volume pinning and the strip width and edge quality are the same for both samples, the dominant pinning contribution should be sought at the film-substrate and film-capping interfaces. The transition temperature $T_\mathrm{c}$ of sample~S is higher than that of sample~T, while the transition widths are nearly identical for both samples, see Fig.~\ref{fig:cooling}(a,\,b). Therefore, possible differences in the flat and smooth MgB$_2$/Au interface (see Fig.\,\ref{fig:structure}) are unlikely to be the origin of the stronger pinning in sample~S. A detailed inspection of the HRTEM image in Fig.~\ref{fig:structure}(a) reveals spatial variations in contrast along the MgB$_2$/MgO interface in sample~S. These variations indicate changes in the electronic---and thus superconducting---properties, with an average lateral periodicity of approximately $6$-$8$\,nm. This length scale is comparable to the coherence length in sample S, implying that the associated pinning potential troughs act individually, placing sample S in the regime of moderately strong pinning.

The HRTEM image in Fig.~\ref{fig:structure}(b) reveals columnar boundaries in the textured film. The presumed enhanced electron scattering at these boundaries is consistent with the nearly twofold higher resistivity of sample~T compared to sample~S. Accordingly, the superconducting order parameter is expected to be strongly reduced at the column boundaries and, in general, this order-parameter variation should give rise to vortex pinning. However, the width of the column walls is much smaller than the superconducting coherence length, which sets the typical radius of a vortex core \cite{Bra95rpp}. Furthermore, the diameters of the columns in Fig.~\ref{fig:structure}(b) range from a relatively uniform $\sim 0.8$\,nm near the Al$_2$O$_3$/MgB$_2$ interface to a more nonuniform distribution of 0.5-4\,nm at a thickness of 20\,nm (corresponding to the MgB$_2$/Au interface in sample T). Thus, in the textured sample each vortex interacts simultaneously with many columns, leading to partial averaging of individual pinning forces and placing sample T closer to the collective pinning regime~\cite{Rui26pms}. This interpretation is supported by fitting the experimental data for sample~T in Fig.~\ref{fig:Arrhenius} to $U_\mathrm{eff}(B) = U_0 B^{-n}$ with $U_0 = 200$\,K and $n = 0.5$, as expected for a 2D collective pinning regime \cite{Kwo16rpp}. In contrast, such a fit is not possible for sample~S.

\subsection{Current-voltage curves}

\begin{figure}
    \centering
    \includegraphics[width=\linewidth]{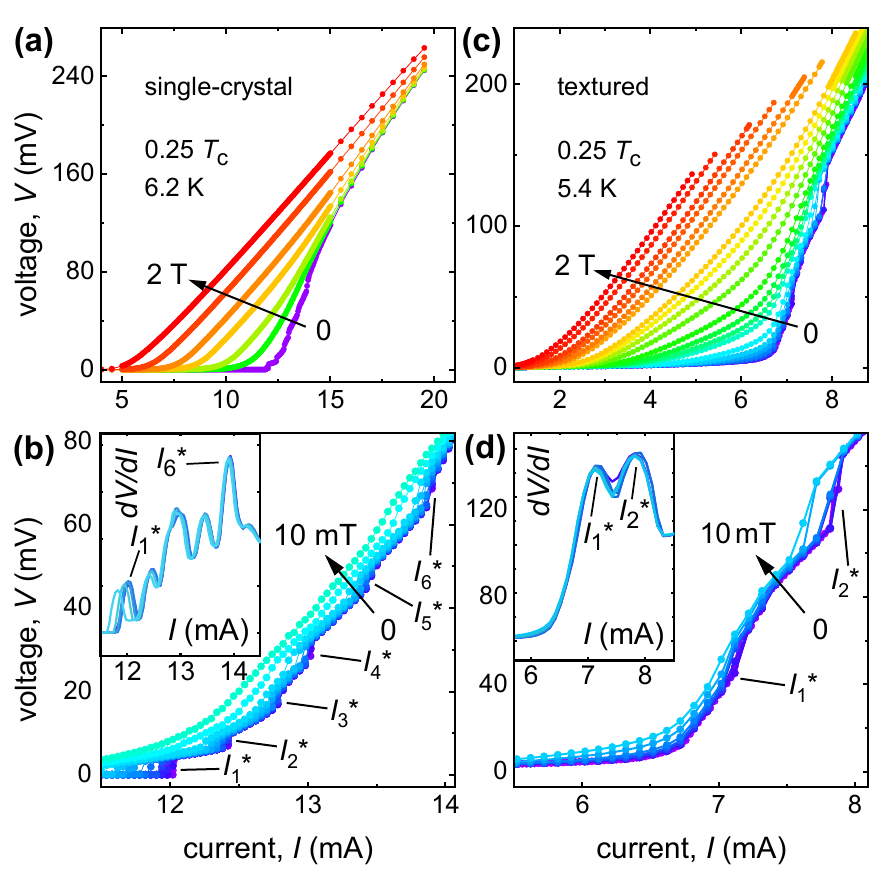}
    \caption{$I$-$V$ curves of the single-crystal (a,\,b) and textured (c,\,d) MgB$_2$ films in magnetic fields ranging from $0$ to $2$\,T. Panels (b, d) show multiple voltage steps (indicated by lines) in the $I$-$V$ curves and the corresponding maxima in the differential resistance (insets) for samples~S and T, respectively, at magnetic fields below $10$\,mT. The currents $I_1^\ast$, $I_2^\ast$, and $I_6^\ast$ in the insets correspond to the voltages which are discussed in the main text.}
    \label{fig:CVC}
\end{figure}

Figure~\ref{fig:CVC} shows the $I$-$V$ curves for both films in magnetic fields up to $2$\,T at $0.25T_\mathrm{c}$. At low fields, the $I$-$V$ curves of sample~S display a zero-voltage plateau with a well-defined critical current which decreases from $I_\mathrm{c} \approx 12$\,mA to about $2$\,mA, with increase of the magnetic field from $0$ to $2$\,T. Here, $I_\mathrm{c}$ is deduced using a voltage criterion of $100$\,nV. The steepness of the $I$-$V$ curves ensures that small variations in the voltage criterion have little effect on the deduced $I_\mathrm{c}$. The resistive transition in sample~T occurs at a factor of two smaller currents, in-line with the low pinning activation energies deduced from the Arrhenius analysis in Fig.\,\ref{fig:Arrhenius}.

At low magnetic fields, the $I$-$V$ curves of both films exhibit multiple jumps in the resistive transition. Specifically, up to six jumps are observed in the $I$-$V$ curve of sample~S in Fig.~\ref{fig:CVC}(b), while two jumps appear in the $I$-$V$ curve of sample~T [Fig.~\ref{fig:CVC}(d)]. These voltage jumps correspond to maxima in the $dV/dI$ curves shown in the insets of Fig.~\ref{fig:CVC}(b,\,d), where a running average over three neighboring points was applied to smooth the derivative data. As the magnetic field increases, the steps in the $I$-$V$ curves become smeared and disappear above $40$\,mT and $50$\,mT for samples S and T, respectively.

\section{Discussion}

\subsection{Multiple voltage transitions}

We first summarize the major experimental findings. First, the textured film exhibits weaker vortex pinning, with the high-resistive state appearing at roughly half the current of the single-crystal film. Second, the $I$-$V$ curves of both samples exhibit multiple voltage transitions at low magnetic fields, with a larger number of steps in the single-crystal film. The origin of the different pinning strengths was already discussed in Sec.~\ref{sec:Arrhenius}. We now turn to a discussion of the multiple voltage transitions.

As the transport current exceeds a depinning threshold, one generally expects a nonlinear transition to a quasi-linear flux-flow regime, characterized by a smooth rise of the $I$-$V$ curve due to fast vortex motion, followed by an abrupt jump to a highly resistive state as a consequence of FFI \cite{Dob24cmp}. Our observations in Fig.\,\ref{fig:CVC} go beyond this simple scenario for spatially uniform systems. Indeed, in the presence of stronger random pinning, the FFI jump to the high-resistive state is generally expected at lower voltages, corresponding to smaller currents and lower vortex velocities compared to the case of weaker pinning~\cite{Sil12njp,Dob19pra}. This earlier onset of FFI arises from a less ordered vortex lattice. The resulting broad distribution of vortex velocities leads to a significant difference between the average vortex velocity, associated with the measured dc voltage, and the maximal vortex velocity within the vortex ensemble,
which is responsible for triggering the FFI \,\cite{Bez19prb}. We note that the current $I^\ast_1$ at which the first jump in the $I$-$V$ curve occurs for sample~S [see the inset of Fig.~\ref{fig:CVC}(b)] depends on the magnetic field, suggesting a possible relation to vortex dynamics. In contrast, the currents corresponding to all subsequent transitions in sample~S, as well as both transitions in sample~T, are much less field-dependent [see the insets in Fig.~\ref{fig:CVC}(c,d)], suggesting a non-vortex origin, or at least a non-global-FFI origin. Furthermore, using formally $v^\ast = V^\ast/(B L)$ to estimate the vortex velocity at the onset of the voltage jumps yields $v^\ast \approx 180$\,m/s for sample~S at $B = 10$\,mT at the first jump $I^\ast_1$ [see Fig.~\ref{fig:CVC}(b)]. At the last jump ($I^\ast_6$), this value increases to $220$\,km/s. For sample~T, the first jump ($I^\ast_1$) gives $v^\ast \approx 140$\,km/s at $10$\,mT while the second jump ($I^\ast_2$) yields  $382$\,km/s. These values are either notably smaller or at least one order of magnitude larger than typical FFI velocities ~\cite{Emb17nac,Dob20nac,Bud22pra}. Thus, one must consider alternative mechanisms for the resistive transition. 

In general, multiple jumps in the $I$-$V$ curves can appear in the presence of phase-slip centers (PSCs) or phase-slip lines (PSLs), the latter being two-dimensional analogues of PSCs \cite{Zol14ltp}. PSCs are typically observed in narrow films with width $W$ on the order of $\xi$, whereas PSLs can occur in wider superconducting films where both $\xi$ and $\lambda$ are smaller than $W$ \cite{Siv03prl}. The oscillations of the superconducting order parameter along a PSL may be nonuniform, propagating as waves---so-called kinematic vortices---that carry the singularities of the order parameter across the film \cite{Sil10prl}. Accordingly, multiple steps in the $I$-$V$ curves in the phase-slip regime are usually observed in superconductors with relatively large $\xi$ and in geometries where several, but not too many, PSCs or PSLs fit between the voltage leads \cite{Zol14ltp}. In practice, for uniform systems, the PSC or PSL mechanism is usually inferred when the distinct linear segments of the $I$-$V$ curve, extrapolated to intersect the current axis, cross at a common point which corresponds to the excess current. This excess current reflects the contribution of the supercurrent that persists even when a finite voltage is present due to phase-slip dynamics. In our study, the strip length of $28\,$µm is likely too long and the coherence length $\xi(T)\simeq10$\,nm is too small for the observation of steps arising solely from PSLs without the involvement of Abrikosov vortices.

The role of Abrikosov vortices in the formation of PSLs can be elucidated as follows. Namely, channels of fast vortex motion in localized regions of the superconducting strip may lead to the formation of PSLs, commonly referred to in this context as vortex rivers~\cite{Sil10prl,Vodolazov2019}. As the transport current increases, the number of vortex rivers grows, resulting in a multi-step $I$-$V$ curve. Previous modeling, based on the generalized TDGL equation coupled with a heat-balance equation, has shown that vortex rivers can evolve into N domains when heat removal from the sample is insufficient \cite{Bez22prb}. The individual resistance steps then correspond to different numbers of N domains, which increase with increasing transport current. The plausibility of attributing the multi-step resistive transitions in both samples to N domains can be further analyzed on the basis of Eq.\,\eqref{eq:jeq}. In this equation, the heat-removal coefficient for our samples is unknown. As an upper-bound estimate, we use the experimentally deduced value $h = 0.27\,$W$\,$K$^{-1}\,$cm$^{-2}$ for a Al$_2$O$_3$/Nb interface \cite{Bez19prb}, which exhibits efficient heat removal. Substituting the experimental data for sample S at $10\,$mT into Eq.\,\eqref{eq:jeq} yields $j_\mathrm{eq} = 409\,$kA/cm$^2$. This value is much smaller than $j^\ast = 6.9\,$MA/cm$^2$ at which the first jump in the $I$-$V$ occurs. The definition of the current(s) $I^\ast$ associated with $j^\ast$ is shown in Fig.\,\ref{fig:CVC}(b). For sample T, $j_\mathrm{eq} = 382\,$kA/cm$^2$ and $j^\ast = 3.9\,$MA/cm$^2$. These estimates place our measurements well within the regime where $j_\mathrm{eq} \ll j^\ast$. The large separation between $j_\mathrm{eq}$ and $j^\ast$ strongly supports a scenario in which normal-domain growth governs the resistive transitions. This condition is naturally satisfied for all other jumps occurring at higher currents in the $I$-$V$ curves of both films. 

\begin{figure*}
    \centering
    \includegraphics[width=\linewidth]{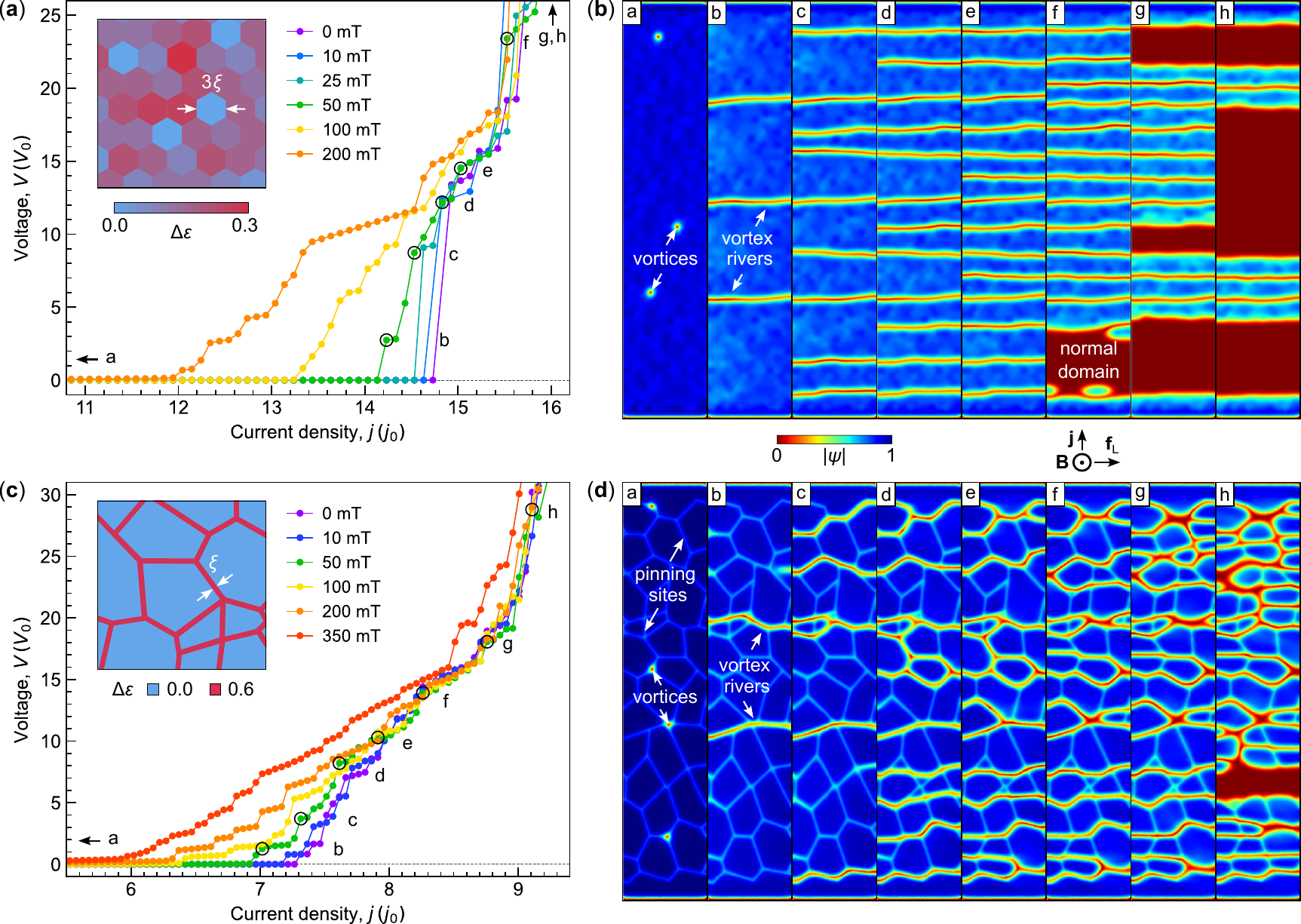}
    \caption{TDGL simulations of the current-driven resistive transition. Panels (a) and (c) show the $I$-$V$ curves, while panels (b) and (d) display snapshots of the superconducting order-parameter magnitude $|\psi|$ at representative points along the corresponding $I$-$V$ curves. The insets in (a) and (c) illustrate the spatial distribution of $|\psi|$ in the simulations, with the parameter $\Delta\epsilon$ controlling the local suppression of the order parameter. The current density $j$ is expressed in units of $j_0 = 4\times 10^{-2}\,\xi B_\mathrm{c2}/(\mu_0 \lambda^2)$, where $B_\mathrm{c2} = \Phi_0/(2\pi\xi^2)$ is the upper critical field, and the voltage $V$ is in units of $V_0 = \xi j_0/\sigma$, with $\sigma$ being the normal-state conductivity.}
    \label{fig:modeling_I-V}
\end{figure*}

\subsection{TDGL modeling}
To elucidate the spatiotemporal dynamics of the superconducting order parameter during the resistive transition, we performed numerical simulations based on the generalized TDGL equation~\cite{Kra78prl,Bis23cpc}. The TDGL equation is valid close to $T_\mathrm{c}$. Far from $T_\mathrm{c}$, the TDGL equation does not reproduce the physics in the vortex core quantitatively, but it still describes the spatiotemporal evolution of vortex matter qualitatively \cite{Ara02rmp}. At low temperatures ($\sim0.25 T_\mathrm{c}$) and low magnetic fields, both gaps in MgB$_2$ are fully open \cite{Dah03prl}, and the material behaves as a single superconducting condensate. Accordingly, we adopted a strong interband-coupling approximation in the TDGL modeling \cite{Dol08prb, Gri16prb}, where fast Cooper-pair tunneling locks the order parameters of all bands, rendering the system effectively single-band for vortex dynamics and allowing a single-band TDGL description.
In two dimensions, the generalized TDGL equation reads
\begin{equation}
    \begin{split}
    \frac{u}{\sqrt{1+\gamma^2|\psi|^2}}&\left(\frac{\partial}{\partial t}+i\varphi+\frac{\gamma^2}{2}\frac{\partial |\psi|^2}{\partial t}\right)\psi\\
    &= (\nabla-i\mathbf{A})^2\psi + (\epsilon-|\psi|^2)\psi
    ,
    \end{split}
    \label{eq:tdgl_psi}
\end{equation}
where $\psi(\mathbf{r},t) = |\psi| e^{i \theta}$ is the superconducting order parameter with the ratio $u=\pi^4/14\zeta(3)\approx5.79$ [$\zeta(x)$: Riemann zeta function] of relaxation times for its amplitude and phase $\theta$ in the dirty-limit regime. The parameter $\gamma=2\tau_\mathrm{E}\Delta_0/\hbar$ involves the inelastic electron-phonon scattering time $\tau_\mathrm{E}$ and the zero-field superconducting gap $\Delta_0$. $\varphi(\mathbf{r}, t)$ is the electric scalar potential and $\mathbf{A}$ the vector potential. $\epsilon(\mathbf{r}) \in [-1, 1]$ is a real-valued parameter that locally modulates the superconducting order parameter and is used to model vortex pinning sites \cite{Kos16prb}.

The total current density satisfies the continuity condition $\nabla\cdot (\mathbf{j}_s + \mathbf{j}_n) = 0$ and involves the supercurrent $\mathbf{j}_s = \mathrm{Im}[\psi^\ast(\nabla-i\mathbf{A})\psi]$ and normal $\mathbf{j}_n = -\nabla\varphi - \partial \mathbf{A}/\partial t$ densities. The electric potential $\varphi(\mathbf{r}, t)$ satisfies the Poisson equation 
    $\nabla^2\varphi = \nabla\cdot\mathrm{Im}[\psi^\ast(\nabla-i\mathbf{A})\psi] - \nabla \cdot \partial \mathbf{A}/\partial t$.
At the strip edges, where vortices enter and exit, superconductor–vacuum boundary conditions are imposed as $\mathbf{n}\cdot(\nabla - i\mathbf{A})\psi = 0$ and $\mathbf{n}\cdot\nabla\varphi = 0$, where $\mathbf{n}$ is the unit vector normal to the interface. At the superconductor–normal-metal interfaces supplying the transport current $j_\mathrm{ext}$, the boundary conditions read $\psi = 0$ and $\mathbf{n}\cdot\nabla\varphi = j_\mathrm{ext}$.

In Eq.\,\eqref{eq:tdgl_psi}, the order-parameter amplitude is normalized to the superconducting carrier density, $|\psi|^2 = n_\mathrm{s}$. Lengths are expressed in units of the coherence length $\xi$, and time is scaled by $\tau_0 = \mu_0 \sigma \lambda^2$, where $\sigma$ is the normal-state conductivity and $\lambda$ the London penetration depth. The vector potential $\mathbf{A}$ is scaled by $\xi B_\mathrm{c2}$. The external current density $j_\mathrm{ext}$ and electric potential $\varphi$ are scaled by $4\xi B_\mathrm{c2}/(\mu_0 \lambda^2)$ and $\xi j_0/\sigma$, respectively. The maximum mesh size was $\xi/2$, and the material parameters are listed in Table~\ref{tab:tdgl_material_params}.  The simulations were performed for a $200\times1000$\,nm$^2$ (width$\times$length) strip in a perpendicular magnetic field. Their results are summarized next.

We first consider the $I$-$V$ curves of a film in which the order parameter varies slightly on a lateral scale of $\sim 3\xi$ [see the inset in Fig.~\ref{fig:modeling_I-V}(a)] and which corresponds to sample S. At subcritical transport currents, the $I$-$V$ curves exhibit a zero-voltage response, with vortices pinned in regions of the strongest order-parameter suppression [regime~a in Fig.~\ref{fig:modeling_I-V}(a,\,b)]. At magnetic fields up to about $300$\,mT, all $I$-$V$ curves display multiple steps, which become progressively smeared at higher fields (not shown). This smearing is attributed to the increased rigidity of the vortex lattice at higher vortex densities~\cite{Bra95rpp}. As the transport current increases, the system evolves through a sequence of dynamic states (denoted b–h), illustrated in Fig.~\ref{fig:modeling_I-V}(a,\,b) for the $I$-$V$ curve at $50$\,mT. Owing to the difference in the widths of the superconducting strip in the model and in the experiment, this field corresponds to approximately $5$\,mT for sample~S in Fig.~\ref{fig:CVC}(a,\,b). 

\begin{table}[t!]
\caption{Material parameters used in the simulations.}
\label{tab:tdgl_material_params}
\small
\begin{center}\begin{tabular}{lcc}
 \hline
 \textbf{Parameter} & \textbf{Denotation} & \textbf{Value}\\
 \hline
 Clean-limit coherence length & $\xi_0$ & 9 nm
 \\
 Electron mean free path & $l$ & 3 nm
 \\
 Coherence length & $\xi(0) = 0.85 \, \sqrt{\xi_0 l}$ & 5 nm
 \\
 London penetration depth & $\lambda(0)$ & 100 nm
 \\
 Inelastic scattering coefficient & $\gamma$ & 10
 \\
 Relative temperature & $T/T_\mathrm{c}$ & 0.3
 \\
 Coherence length & $\xi=\frac{\xi(0)}{\sqrt{1-T/T_\mathrm{c}}}$ & 5\,nm
 \\
 Penetration depth & $\lambda= \frac{0.615 \, \lambda(0)\sqrt{\xi_0}}{\sqrt{l (1-T/T_\mathrm{c})}}$ & 127\,nm\\
\hline
\end{tabular}\end{center}
\end{table}
The simulations suggest that the first voltage step is associated with the formation of one or more channels of fast-moving vortices [regime~b in Fig.~\ref{fig:modeling_I-V}(a,\,b)]. At such high velocities, these vortex chains form vortex rivers (PSLs)~\cite{Zol14ltp,Siv03prl}. Subsequent voltage transitions correspond to states with an increasing number of vortex rivers [regimes~c–e in Fig.~\ref{fig:modeling_I-V}(a,\,b)]. With further increase of the transport current, heat removal to the substrate becomes insufficient, leading to the nucleation of an N domain within the strip. As $j$ exceeds the $j_\mathrm{eq}$, this domain expands (regime~f), followed by the formation and growth of additional domains (regime~g), until the entire strip transitions to the normal state [regime~h in Fig.~\ref{fig:modeling_I-V}(a,\,b)].

We now proceed to the analysis of the $I$-$V$ curves for the film with a network of channels of width $\sim\xi$ with suppressed order parameter, as illustrated in the inset in Fig.~\ref{fig:modeling_I-V}(c) and which corresponds to sample T. While the net pinning in this case is collective and weak, we assume that there might be some areas which allow for vortex channeling. Simultaneously, on the basis of structural characterization of film T we assume that the density of such channels is rather small. The simulated $I$-$V$ curve exhibits a zero-voltage response up to the depinning currents which are a factor of two smaller than those for sample S. In this regime a in Fig.~\ref{fig:modeling_I-V}(c,\,d) the vortices are pinned to the nodes of the pinning sites network. 

When the transport current exceeds the depinning threshold, a low-resistive regime emerges, characterized by a slightly S-shaped global $I$-$V$ response with superimposed small voltage steps. Because overheating is neglected in the TDGL simulations, these steps are expected to be smeared in the experimental $I$-$V$ curves. The corresponding vortex configurations evolve from a single vortex river to multiple rivers with increasing current [regimes~b-g in Fig.~\ref{fig:modeling_I-V}(c,\,d)], with the texture stabilizing these channels by providing multiple pathways. 
In this way, the TDGL simulations reproduce the essential features of the experimental $I$-$V$ curves and indicate that the multiple voltage transitions are associated with distinct dynamical regimes of vortex rivers and normal domains in the superconducting strip.

Finally, we recall that the TDGL modeling was performed within the strong interband-coupling approximation~\cite{Dol08prb,Gri16prb}. In MgB$_2$, the $\sigma$ and $\pi$ bands possess distinct superconducting parameters and different sensitivities to structural defects. The $\sigma$ band, characterized by strong electron-phonon coupling, is primarily affected by large structural defects. In contrast, the $\pi$ band, with weaker electron–phonon coupling and a smaller superconducting gap, is more sensitive to disorder \cite{Kas18aip}. Consequently, structural defects have a more pronounced impact on the $\pi$ band, where scattering more strongly perturbs superconductivity. Extending the modeling to explicitly account for the two-gap nature of MgB$_2$ is therefore expected to yield a broader distribution of pinning strengths, as well as a richer variety of vortex pathways and coexisting dynamical states.

\subsection{Impact of heat removal}
We note that our TDGL simulations were performed without accounting for heat removal, which would require solving the TDGL equations in conjunction with the heat equation. Accounting for a finite heat-removal rate is expected to (i) reduce the current range over which vortex rivers persist in favor of regimes dominated by N domains, (ii) accelerate the growth of N domains, and (iii) promote the smearing of small voltage steps.

At the same time, the presence of multiple steps in the $I$-$V$ curve of both films indicates that superconductivity is not destroyed throughout the entire sample simultaneously. Instead, several distinct dynamic states exist, each with a different resistance. The dissipation generated in these states does not drive the sample into the normal state, indicating that it is well balanced by heat removal. The greater number of voltage transitions in sample~S, occurring at currents roughly twice as high as in sample~T, indicates more efficient heat removal in sample~S. This can be attributed to the epitaxial nature of the stack Al$_2$O$_3$/MgO/MgB$_2$ which facilitates the transmission of phonons into the substrate. On the contrary, the large lattice mismatch at the Al$_2$O$_3$/MgB$_2$ interface and the 1.5\,nm-thick interfacial amorphous MgAlBO$_\mathrm{x}$ layer for the textured film in Fig.~\ref{fig:structure}(b), likely results in acoustic impedance mismatch, i.e., a high Kapitza thermal boundary resistance~\cite{Kap41jph,Pol69rmp}. This barrier impedes heat transfer, reflecting nonequilibrium phonons back into the film and creating a phonon bottleneck~\cite{Sid18prb}, which accelerates the transition to the normal state. 

\section{Conclusion}
In summary, we have investigated the influence of two types of interfacial and volumetric crystal defects on the resistive transition in 20\,nm-thick MgB$_2$ films. Interfacial roughness leads to stronger pinning, as evidenced by a higher activation energy compared to the weaker defects associated with slight misalignment between columns in textured films. As a result, the current at which resistance first becomes finite is higher in the single-crystal film than in the textured one. Analysis of the $I$-$V$ characteristics reveals multiple voltage jumps, for which a conventional global flux-flow instability can be ruled out, as it would require unrealistically high vortex velocities. Instead, comparison with time-dependent Ginzburg–Landau simulations shows that the resistive transition is governed by the nucleation and growth of normal domains. While both types of defects have a comparable impact on the emergence of step-like transitions, the efficiency of heat removal at the film-buffer interface becomes the decisive factor in stabilizing or suppressing these steps. The single-crystal film sustains significantly higher currents, consistent with more efficient heat removal, which we attribute to improved acoustic matching at the film-buffer interface. These findings demonstrate that both the intrinsic film structure and the properties of the film–buffer interface critically determine vortex dynamics and the low-dissipative current-carrying capacity at low temperatures. More broadly, our results offer insights for optimizing superconducting devices based on MgB$_2$, where precise control of dissipation at high transport currents is required.

\begin{acknowledgments}
C.S. acknowledges financial support by the Vienna Doctoral School in Physics (VDSP). This work was funded by the Austrian Science Fund (FWF), Grant No. I 6079 (FluMag). The work of A.P. was funded by the Deutsche Forschungsgemeinschaft (DFG, German Research Foundation) under Germany's Excellence Strategy -- EXC-2123 QuantumFrontiers -- 390837967, project Q-53. A.P. gratefully acknowledges the use of the CryoCore and CryoCube simulation workstations at CryoQuant/TU Braunschweig and thanks for the support of the Braunschweig International Graduate School of Metrology (B-IGSM). T.H. thanks CCDaum, CC3M and CCMinalor at Institut Jean Lamour for their work on UHV growth, HRTEM imaging and on film patterning  respectively. This work was supported by the French National Research Agency through the France 2030 government grants ``PEPR-SPIN'' ANR-24-EXSP-0012 and ``Lorraine Initiative of Excellence'' ANR-15-IDEX-04-LUE. This research is based upon work from COST Action CA21144 (SuperQuMap) supported by the European Cooperation in Science and Technology. For the purpose of open access, the authors have applied a CC BY public copyright license to any Author Accepted Manuscript version arising from this submission.
\end{acknowledgments}

\section*{Data Availability}
The experimental data supporting the findings of this article are openly available~\cite{dataset}.

\bibliography{./SupercondTheory, ./MgB2, ./SNSPDs, ./fluxonics}

@Article{Nag01nat,
  author     = {Jun Nagamatsu and Norimasa Nakagawa and Takahiro Muranaka and Yuji Zenitani and Jun Akimitsu},
  journal    = {Nature},
  title      = {Superconductivity at {39 K} in magnesium diboride},
  year       = {2001},
  issn       = {0028-0836},
  number     = {6824},
  pages      = {63-64},
  volume     = {410},
  comment    = {First discovery of superconductivity properties in bulk MgB2 at high temperatures of 39 K},
  doi        = {10.1038/35065039},
  publisher  = {Springer Science and Business Media LLC},
  ranking    = {rank4},
  readstatus = {skimmed},
}

@Article{Fer01sst,
  author     = {Ferdeghini, C and Ferrando, V and Grassano, G and Ramadan, W and Bellingeri, E and Braccini, V and Marré, D and Manfrinetti, P and Palenzona, A and Borgatti, F and Felici, R and Lee, T-L},
  journal    = {Supercond. Sci. Technol.},
  title      = {Growth of c-oriented {MgB$_2$} thin films by pulsed laser deposition: structural characterization and electronic anisotropy},
  year       = {2001},
  issn       = {1361-6668},
  month      = oct,
  number     = {11},
  pages      = {952--957},
  volume     = {14},
  comment    = {Early paper on MgB2 thin films using Sapphire as substrate, fabricated by pulsed laser deposition 

Finds rather high critical temperature close to 38 K},
  doi        = {10.1088/0953-2048/14/11/311},
  publisher  = {IOP Publishing},
  ranking    = {rank3},
  readstatus = {skimmed},
}

@Article{Zha13tas,
  author     = {Chen Zhang and Yue Wang and Da Wang and Yan Zhang and Qing-Rong Feng and Zi-Zhao Gan},
  journal    = {IEEE Trans. Appl. Supercond.},
  title      = {Hybrid Physical–Chemical Vapor Deposition of Ultrathin $\hbox{MgB}_{2}$ Films on {MgO} Substrate With High {$T_{c}$} and {$J_{c}$}},
  year       = {2013},
  issn       = {1051-8223},
  number     = {3},
  pages      = {7500204-7500204},
  volume     = {23},
  comment    = {MgB2 thin film fabrication with single photon detectors in mind

Achieve high critical temperature in thin films and high current density},
  doi        = {10.1109/tasc.2012.2230212},
  publisher  = {Institute of Electrical and Electronics Engineers (IEEE)},
  ranking    = {rank4},
  readstatus = {skimmed},
}

@Article{Tian2002,
  author    = {Tian, W. and Pan, X. Q. and Bu, S. D. and Kim, D. M. and Choi, J. H. and Patnaik, S. and Eom, C. B.},
  journal   = {Appl. Phys. Lett.},
  title     = {Interfacial structure of epitaxial {MgB}$_{2}$ thin films grown on (0001) sapphire},
  year      = {2002},
  issn      = {1077-3118},
  month     = jul,
  number    = {4},
  pages     = {685--687},
  volume    = {81},
  comment   = {Study on the interfacial structure of MgB2 that is grown by MBE on a sapphire substrate, showing formation of textured surfaces},
  doi       = {10.1063/1.1489101},
  publisher = {AIP Publishing},
  ranking   = {rank4},
}

@Article{Saito2004,
  author    = {Saito, A. and Shimakage, H. and Kawakami, A. and Wang, Z. and Kuroda, K. and Abe, H. and Naito, M. and Moon, W.J. and Kaneko, K. and Mukaida, M. and Ohshima, S.},
  journal   = {Physica C},
  title     = {{XRD and TEM} studies of as-grown {MgB}$_{2}$ thin films deposited on r- and c-plane sapphire substrates},
  year      = {2004},
  issn      = {0921-4534},
  month     = oct,
  pages     = {1366--1370},
  volume    = {412–414},
  comment   = {Studies of growth of {MgB}$_{2}$ on sapphire substrate. Compares different crystal structures of the sapphire showing some with stronger alloy formation},
  doi       = {10.1016/j.physc.2003.12.100},
  publisher = {Elsevier BV},
  ranking   = {rank3},
}

@Article{Xi2009,
  author    = {Xi, X X},
  journal   = {Supercond. Sci. Technol.},
  title     = {{MgB}$_{2}$ thin films},
  year      = {2009},
  issn      = {1361-6668},
  month     = mar,
  number    = {4},
  pages     = {043001},
  volume    = {22},
  comment   = {Review on growth and fabrication techniques for {MgB}$_{2}$ and some application ideas},
  doi       = {10.1088/0953-2048/22/4/043001},
  publisher = {IOP Publishing},
  ranking   = {rank3},
}

@Article{Ueda2003,
  author    = {Ueda, K. and Naito, M.},
  journal   = {J. Appl. Phys.},
  title     = {In situ growth of superconducting {MgB}$_{2}$ thin films by molecular-beam epitaxy},
  year      = {2003},
  issn      = {1089-7550},
  month     = feb,
  number    = {4},
  pages     = {2113--2120},
  volume    = {93},
  comment   = {Fabrication and growth of MgB2 via molecular beam epitaxy (MBE)},
  doi       = {10.1063/1.1537460},
  publisher = {AIP Publishing},
  ranking   = {rank4},
}

@Article{Li2017,
  author    = {Li, Lin and Zhang, Hui and Yang, Yi‐Hang and Miao, Guo‐Xing},
  journal   = {Adv. Eng. Mater.},
  title     = {High‐Quality Epitaxial {MgB}$_{2}$  {J}osephson Junctions Grown by Molecular Beam Epitaxy},
  year      = {2017},
  issn      = {1527-2648},
  month     = feb,
  number    = {5},
  volume    = {19},
  pages = {1600792},
  comment   = {High quality MgB2 films and Josephson junctions grown via molecular beam epitaxy (MBE)},
  doi       = {10.1002/adem.201600792},
  publisher = {Wiley},
  ranking   = {rank3},
}

@Article{Jo2002,
  author    = {Jo, W. and Huh, J-U. and Ohnishi, T. and Marshall, A. F. and Beasley, M. R. and Hammond, R. H.},
  journal   = {Appl. Phys. Lett.},
  title     = {In situ growth of superconducting {MgB}$_{2}$ thin films with preferential orientation by molecular-beam epitaxy},
  year      = {2002},
  issn      = {1077-3118},
  month     = may,
  number    = {19},
  pages     = {3563--3565},
  volume    = {80},
  comment   = {MgB2 grown via molecular beam epitaxy (MBE) with particular orientation

Shows difference of single crystal vs. textured structures},
  doi       = {10.1063/1.1478151},
  publisher = {AIP Publishing},
  ranking   = {rank4},
}

@Article{Gu2005,
  author    = {Gu, Lin and Moeckly, Brian H. and Smith, David J.},
  journal   = {J. Cryst. Growth},
  title     = {Electron microscopy studies of epitaxial {MgB}$_{2}$ superconducting thin films grown by in situ reactive evaporation},
  year      = {2005},
  issn      = {0022-0248},
  month     = jul,
  number    = {3–4},
  pages     = {602--611},
  volume    = {280},
  comment   = {Electron microscope observations of MgB2 films grown on different substrates, showing interface layer (textured) and single crystal structures},
  doi       = {10.1016/j.jcrysgro.2005.03.066},
  publisher = {Elsevier BV},
  ranking   = {rank4},
}

@Article{Jorgensen2001,
  author    = {Jorgensen, J. D. and Hinks, D. G. and Short, S.},
  journal   = {Phys. Rev. B},
  title     = {Lattice properties of {MgB}$_{2}$ versus temperature and pressure},
  year      = {2001},
  issn      = {1095-3795},
  month     = may,
  number    = {22},
  pages     = {224522},
  volume    = {63},
  comment   = {Determination of c lattice parameter in MgB2},
  doi       = {10.1103/physrevb.63.224522},
  publisher = {American Physical Society (APS)},
  ranking   = {rank3},
}

@article{Kra78prl,
title = {Theory of Dissipative Current-Carrying States in Superconducting Filaments},
author = {
Kramer, L. and 
Watts-Tobin, R. J.},
j1 = {PRL},
journal2 = {Physical Review Letters},
journal = {Phys. Rev. Lett.},
publisher = {American Physical Society},
volume = {40},
number = {15},
year = {1978},
month = {04},
pages = {1041--1044},
doi = {10.1103/PhysRevLett.40.1041},

date = {1978/04/10/}
}

@article{Bis23cpc,
title = {{pyTDGL}: {Time-dependent Ginzburg-Landau} in {Python}},
author = {Bishop-Van Horn, Logan},
journal2 = {Computer Physics Communications},
journal = {Comput. Phys. Commun.},
isbn = {0010-4655},
volume = {291},
year = {2023},
pages = {108799},
doi = {10.1016/j.cpc.2023.108799},

date = {2023/10/01/}
}

@article{Kos16prb,
  title = {Optimization of vortex pinning by nanoparticles using simulations of the time-dependent {G}inzburg-{L}andau model},
  author = {Koshelev, A. E. and Sadovskyy, I. A. and Phillips, C. L. and Glatz, A.},
  journal = {Phys. Rev. B},
  volume = {93},
  issue = {6},
  pages = {060508},
  numpages = {5},
  year = {2016},
  month = {Feb},
  publisher = {American Physical Society},
  doi = {10.1103/PhysRevB.93.060508},
  url = {https://link.aps.org/doi/10.1103/PhysRevB.93.060508}
}

@misc{dataset,
    author = {Schmid, Clemens},
    year = {2026}, 
    title = {Research data for the research article ``{Crystal} structure effects on vortex dynamics in superconducting {MgB$_2$} thin films''}, 
    publisher = {Mendeley Data},
    note={{M}endeley {D}ata, {DOI}: \href{https://doi.org/10.17632/42k2ym76s7.1}{10.17632/42k2ym76s7.1}}
}

@Article{Vodolazov2019,
  author    = {D. Yu Vodolazov},
  journal   = {Superconductor Science and Technology},
  title     = {Flux-flow instability in a strongly disordered superconducting strip with an edge barrier for vortex entry},
  year      = {2019},
  issn      = {0953-2048},
  number    = {11},
  pages     = {115013},
  volume    = {32},
  doi       = {10.1088/1361-6668/ab4168},
  publisher = {IOP Publishing},
}

@Article{Tsukui2001,
  author     = {S. Tsukui and M. Adachi and R. Oshima and H. Nakajima and F. Toujou and K. Tsukamoto and T. Tabata},
  journal    = {Physica C},
  title      = {Oxygen tracer diffusion in the {YBa$_2$Cu$_3$O$_y$} superconductor},
  year       = {2001},
  issn       = {0921-4534},
  number     = {4},
  pages      = {357-362},
  volume     = {351},
  comment    = {Example of Arrhenius plot with two differnt slopes in YBCO for determination of phase transition temperature},
  doi        = {10.1016/s0921-4534(00)01640-3},
  publisher  = {Elsevier BV},
  ranking    = {rank3},
  readstatus = {skimmed},
}

@Article{Dobrovolskiy,
  author     = {O. V. Dobrovolskiy and E. Begun and M. Huth and V. A. Shklovskij},
  journal    = {New J. Phys.},
  title      = {Electrical transport and pinning properties of {Nb} thin films patterned with focused ion beam-milled washboard nanostructures},
  year       = {2012},
  issn       = {1367-2630},
  number     = {11},
  pages      = {113027},
  volume     = {14},
  comment    = {Magneto-transport properties of FIB-milled Nb film, pinning potential landscape

Quantifies the activation energies of vortex lattice parameters using Arrhenius plots and Arrhenius analysis},
  doi        = {10.1088/1367-2630/14/11/113027},
  publisher  = {IOP Publishing},
  ranking    = {rank3},
  readstatus = {skimmed},
}

@Article{Gre25apl,
author={Greenfield, J.
and Bell, C.
and Faramarzi, F.
and Kim, C.
and Basu Thakur, R.
and Wandui, A.
and Frez, C.
and Mauskopf, P.
and Cunnane, D.},
title={Kinetic inductance and non-linearity of {MgB}$_{2}$ films at 4\,{K}},
journal={Appl. Phys. Lett.},
year={2025},
month={Jan},
day={16},
volume={126},
number={2},
pages={022602},
doi={10.1063/5.0245866},
url={https://doi.org/10.1063/5.0245866}
}

@Article{Cha24nac,
author={Charaev, Ilya
and Batson, Emma K.
and Cherednichenko, Sergey
and Reidy, Kate
and Drakinskiy, Vladimir
and Yu, Yang
and Lara-Avila, Samuel
and Thomsen, Joachim D.
and Colangelo, Marco
and Incalza, Francesca
and Ilin, Konstantin
and Schilling, Andreas
and Berggren, Karl K.},
title={Single-photon detection using large-scale high-temperature {MgB}$_{2}$ sensors at {20 K}},
journal={Nat. Commun.},
year={2024},
month={May},
day={10},
volume={15},
number={1},
pages={3973},
doi={10.1038/s41467-024-47353-x},
url={https://doi.org/10.1038/s41467-024-47353-x}
}

@ARTICLE{Sil12njp,
  author = {Silhanek, A. V. and Leo, A. and Grimaldi, G. and Berdiyorov, G. R.
	and Milosevic, M. V and Nigro, A. and Pace, S. and Verellen, N. and
	Gillijns, W. and Metlushko, V. and Ili\'c, B. and Zhu, X. and Moshchalkov,
	V. V.},
  title = {Influence of artificial pinning on vortex lattice instability in
	superconducting films},
  journal = {New J. Phys.},
  year = {2012},
  volume = {14},
  pages = {053006},
  number = {5},
  url = {http://stacks.iop.org/1367-2630/14/i=5/a=053006}
}

@ARTICLE{Bez84ltp,
  author = {Bezuglyj, A. I. and Shklovskij, V. A.},
  title = {Thermal domains in inhomogeneous current-carrying superconductors.
	{C}urrent-voltage characteristics and dynamics of domain formation
	after current jumps},
  journal = {J. Low Temp. Phys.},
  year = {1984},
  volume = {57},
  pages = {227--247},
  number = {3},
  month = {Nov},
  day = {01},
  doi = {10.1007/BF00681190},
  issn = {1573-7357},
  url = {https://doi.org/10.1007/BF00681190}
}

@ARTICLE{Gur84spu,
  author = {Gurevich, A. V. and Mints, R. G.},
  journal = {Sov. Phys. Usp.},
  year = {1984},
  volume = {27},
  pages = {19}
}

@article{Wat81ltp,
  title        = {Nonequilibrium theory of dirty, current‑carrying superconductors: phase‑slip oscillators in narrow filaments near ${T_c}$},
  author       = {Watts‑Tobin, R.~J. and Kr{\"a}henb{\"u}hl, Y. and Kramer, L.},
  journal      = {J. Low Temp. Phys.},
  volume       = {42},
  number       = {5},
  pages        = {459--501},
  year         = {1981},
  publisher    = {Springer},
  doi          = {10.1007/BF00117427},
  url          = {https://doi.org/10.1007/BF00117427}
}

@Article{Ivl84phb,
author={Ivlev, B. I.
and Kopnin, N. B.},
title={Dynamical processes in current-carrying superconductors},
journal={Physica B+C},
year={1984},
month={Nov},
day={01},
volume={126},
number={1},
pages={346-353},
url={https://www.sciencedirect.com/science/article/pii/0378436384901876}
}

@ARTICLE{Ada15prb,
  author = {Adami, O.-A. and Jelic, Z. L. and Xue, C. and Abdel-Hafiez, M. and
	Hackens, B. and Moshchalkov, V. V. and Milosevic, M. V. and Van de
	Vondel, J. and Silhanek, A. V.},
  title = {Onset, evolution, and magnetic braking of vortex lattice instabilities
	in nanostructured superconducting films},
  journal = {Phys. Rev. B},
  year = {2015},
  volume = {92},
  pages = {134506},
  month = {Oct},
  doi = {10.1103/PhysRevB.92.134506},
  issue = {13},
  numpages = {9},
  publisher = {American Physical Society},
  url = {https://link.aps.org/doi/10.1103/PhysRevB.92.134506}
}

@Article{Buh15nac,
author={Buh, Jo{\v{z}}e
and Kabanov, Viktor
and Baranov, Vladimir
and Mrzel, Ale{\v{s}}
and Kovi{\v{c}}, Andrej
and Mihailovic, Dragan},
title={Control of switching between metastable superconducting states in $\delta$-{MoN} nanowires},
journal={Nat. Commun.},
year={2015},
month={Dec},
day={21},
volume={6},
number={1},
pages={10250},
doi={10.1038/ncomms10250},
url={https://doi.org/10.1038/ncomms10250}
}

@BOOK{Bar82boo,
  title = {Physics and Applications of the Josephson Effect},
  publisher = {John Wiley \& Sons, New York},
  year = {1982},
  author = {A. Barone and G. Patterno}
}

@article{Siv03prl,
  title = {Josephson Behavior of Phase-Slip Lines in Wide Superconducting Strips},
  author = {Sivakov, A. G. and Glukhov, A. M. and Omelyanchouk, A. N. and Koval, Y. and M\"uller, P. and Ustinov, A. V.},
  journal = {Phys. Rev. Lett.},
  volume = {91},
  issue = {26},
  pages = {267001},
  numpages = {4},
  year = {2003},
  month = {Dec},
  publisher = {American Physical Society},
  doi = {10.1103/PhysRevLett.91.267001},
  url = {https://link.aps.org/doi/10.1103/PhysRevLett.91.267001}
}

@article{Rey02prb,
  title = {Current-induced highly dissipative domains in high-${T}_{c}$ thin films},
  author = {Reymond, S. and Antognazza, L. and Decroux, M. and Koller, E. and Reinert, P. and Fischer, \O{}.},
  journal = {Phys. Rev. B},
  volume = {66},
  issue = {1},
  pages = {014522},
  numpages = {7},
  year = {2002},
  month = {Jul},
  publisher = {American Physical Society},
  doi = {10.1103/PhysRevB.66.014522},
  url = {https://link.aps.org/doi/10.1103/PhysRevB.66.014522}
}

@article{Sil10prl,
  title        = {Formation of stripelike flux patterns obtained by freezing kinematic vortices in a superconducting {Pb} film},
  author       = {Silhanek, A. V. and Milo\v{s}evi\'{c}, M. V. and Kramer, R. B. G. and Berdiyorov, G. R. and Van de Vondel, J. and Luccas, R. F. and Puig, T. and Peeters, F. M. and Moshchalkov, V. V.},
  journal      = {Phys. Rev. Lett.},
  volume       = {104},
  number       = {1},
  pages        = {017001},
  year         = {2010},
  publisher    = {American Physical Society},
  doi          = {10.1103/PhysRevLett.104.017001},
  url          = {https://doi.org/10.1103/PhysRevLett.104.017001}
}

@Article{Cat25nac,
author={Cattaneo, Roger
and Efimov, Artemii E.
and Shiianov, Kirill I.
and Kieler, Oliver
and Krasnov, Vladimir M.},
title={Cascade switching current detectors based on arrays of {Josephson} junctions},
journal={Nat. Commun.},
year={2025},
month={Aug},
day={25},
volume={16},
number={1},
pages={7927},
issn={2041-1723},
doi={10.1038/s41467-025-63360-y},
url={https://doi.org/10.1038/s41467-025-63360-y}
}

@Article{Gol01apl,
author={Gol'tsman, G. N.
and Okunev, O.
and Chulkova, G.
and Lipatov, A.
and Semenov, A.
and Smirnov, K.
and Voronov, B.
and Dzardanov, A.
and Williams, C.
and Sobolewski, Roman},
title={Picosecond superconducting single-photon optical detector},
journal={Appl. Phys. Lett.},
year={2001},
month={Aug},
day={06},
volume={79},
number={6},
pages={705-707},
issn={0003-6951},
doi={10.1063/1.1388868},
url={https://doi.org/10.1063/1.1388868}
}

@article{Sil25apr,
  author  = {A. V. Silhanek and L. Jiang and C. Xue and Benoît Vanderheyden},
  title   = {Impact of border defects on the magnetic flux penetration in superconducting films},
  journal = {Appl. Phys. Rev.},
  year    = {2025},
  volume  = {12},
  number  = {4},
  pages   = {041324},
  month   = {Dec},
  doi     = {10.1063/5.0282694},
  url     = {https://pubs.aip.org/apr/article/12/4/041324/3374475/Impact-of-border-defects-on-the-magnetic-flux}
}

@article{Kwo16rpp,
  author       = {Kwok, Wai-Kwong and Welp, Ulrich and Glatz, Andreas and Koshelev, Alexei E. and Kihlstrom, Karen J. and Crabtree, George W.},
  title        = {Vortices in high-performance high-temperature superconductors},
  journal      = {Rep. Progr. Phys.},
  volume       = {79},
  number       = {11},
  pages        = {116501},
  year         = {2016},
  doi          = {10.1088/0034-4885/79/11/116501}
}

@article{Vod17pra,
  author  = {D. Yu. Vodolazov},
  title   = {Photon detection by superconducting nanowires: A time-dependent {Ginzburg-Landau} approach},
  journal = {Phys. Rev. Appl.},
  volume  = {7},
  pages   = {034014},
  year    = {2017},
  doi     = {10.1103/PhysRevApplied.7.034014}
}

@article{Bud22pra,
title = {Rising speed limits for fluxons via edge-quality improvement in wide {M}o{S}i thin films},
author = {Budinsk\'a \emph{et al.}, B.},
journal = {Phys. Rev. Appl.},
volume = {17},
issue = {3},
pages = {034072},
numpages = {12},
year = {2022},
publisher = {American Physical Society},
doi = {10.1103/PhysRevApplied.17.034072},
url = {https://link.aps.org/doi/10.1103/PhysRevApplied.17.034072},
}

@Inbook{Dob24cmp,
author={Dobrovolskiy, Oleksandr},
title={Fast dynamics of vortices in superconductors},
bookTitle={Encycl. Cond. Matt. Phys.},
year={2024},
day={01},
publisher={Elsevier},
address={Amsterdam},
pages={735},
url={https://www.sciencedirect.com/science/article/pii/B9780323908009000159},
doi = {10.1016/B978-0-323-90800-9.00015-9},
}

@ARTICLE{Bez22prb,
author = {Bezuglyj, A. and Shklovskij, V. and Budinsk\'a, B. and Aichner,
B. and Bevz, V. and Mikhailov, M. and Vodolazov, D. and
Lang, W. and Dobrovolskiy, Oleksandr},
title = {Vortex jets generated by edge defects in current-carrying superconductor
thin strips},
journal = {Phys. Rev. B},
year = {2022},
volume = {105},
pages = {214507},
doi = {10.1103/PhysRevB.105.214507},
issue = {21},
numpages = {12},
publisher = {American Physical Society},
url = {https://link.aps.org/doi/10.1103/PhysRevB.105.214507},
}

@ARTICLE{Dob20nac,
  author = {Dobrovolskiy, Oleksandr and Vodolazov, D. and Porrati, F. and Sachser, 	R. and Bevz, V. and Mikhailov, M. and Chumak, A. and Huth, 	M.},
  title = {Ultra-fast vortex motion in a direct-write {Nb-C} superconductor},
  journal = {Nat. Commun.},
  year = {2020},
  volume = {11},
  pages = {3291},
  number = {1},
  day = {03},
  doi = {10.1038/s41467-020-16987-y},
  issn = {2041-1723},
  url = {https://doi.org/10.1038/s41467-020-16987-y}
}

@Article{Dob25nan,
author={Dobrovolskiy, Oleksandr
and Wang, Q.
and Vodolazov, D.
and Sachser, R.
and Huth, M.
and Knauer, S.
and Buzdin, A.},
title={Moving {Abrikosov} vortex lattices generate sub-40-nm magnons},
journal={Nat. Nanotechn.},
year={2025},
month={Oct},
day={16},
volume={20},
pages={1764},
issn={1748-3395},
doi={10.1038/s41565-025-02024-w},
url={https://doi.org/10.1038/s41565-025-02024-w}
}

@article{Bev23pra,
  author       = {V. Bevz and M. Mikhailov and B. Budinská and S. Lamb‑Camarena and S. Shpilinska and A. Chumak and M. Urbánek and M. Arndt and W. Lang and O. Dobrovolskiy},
  title        = {Vortex Counting and Velocimetry for Slitted Superconducting Thin Strips},
  journal      = {Phys. Rev. Appl.},
  volume       = {19},
  number       = {3},
  pages        = {034098},
  year         = {2023},
  doi          = {10.1103/PhysRevApplied.19.034098},
}

@ARTICLE{Kap41jph,
  author = {P. L. Kapitza},
  journal = {J. Phys.},
  year = {1941},
  volume = {4},
  pages = {181}
}

@Article{Sza02pcs,
author={Szab{\'o}, P.
and Samuely, P.
and Jansen, A. G. M.
and Klein, T.
and Marcus, J.
and Fruchart, D.
and Miraglia, S.},
title={Magnetotransport and the upper critical magnetic field in {MgB2}},
journal={Physica C},
year={2002},
month={Mar},
day={15},
volume={369},
number={1},
pages={250-253},
url={https://www.sciencedirect.com/science/article/pii/S0921453401012527}
}

@article{Gri16prb,
  title = {Effective {G}inzburg-{L}andau free energy functional for multi-band isotropic superconductors},
  author = {Konstantin V. Grigorishin},
  journal = {Phys. Lett. A},
  volume = {380},
  number = {20},
  pages = {1781-1787},
  year = {2016},
  issn = {0375-9601},
  doi = {10.1016/j.physleta.2016.03.023},

  langid = {english},
}

@ARTICLE{Loe19acs,
  author = {L{\"o}sch, S. and Alfonsov, A. and Dobrovolskiy, O. V. and Keil,
	R. and Engemaier, V. and Baunack, S. and Li, G. and Schmidt, O. G.
	and B{\"u}rger, D.},
  title = {Microwave Radiation Detection with an Ultra-Thin Free-Standing Superconducting
	Niobium Nanohelix},
  journal = {ACS Nano},
  year = {2019},
  volume = {13},
  pages = {2948},
  day = {04},
  doi = {10.1021/acsnano.8b07280},
  issn = {1936-0851},
  publisher = {American Chemical Society},
  url = {https://doi.org/10.1021/acsnano.8b07280}
}

@article{Dol08prb,
  title = {Strong electron-phonon interaction in multiband superconductors},
  author = {
    Dolgov, O. V. and 
    Golubov, A. A.},
  journal = {Phys. Rev. B},
  volume = {77},
  issue = {21},
  pages = {214526},
  numpages = {5},
  year = {2008},
  month = {Jun},
  publisher = {American Physical Society},
  doi = {10.1103/PhysRevB.77.214526},

  langid = {english},
}

@article{Ara02rmp,
  title = {The world of the complex {Ginzburg-Landau} equation},
  author = {Aranson, Igor S. and Kramer, Lorenz},
  journal = {Rev. Mod. Phys.},
  volume = {74},
  issue = {1},
  pages = {99--143},
  numpages = {0},
  year = {2002},
  month = {Feb},
  publisher = {American Physical Society},
  doi = {10.1103/RevModPhys.74.99},
  url = {https://link.aps.org/doi/10.1103/RevModPhys.74.99}
}

@article{Dah03prl,
  title = {Fermi Surface Topology and the Upper Critical Field in Two-Band Superconductors: Application to ${\mathrm{M}\mathrm{g}\mathrm{B}}_{2}$},
  author = {Dahm, T. and Schopohl, N.},
  journal = {Phys. Rev. Lett.},
  volume = {91},
  issue = {1},
  pages = {017001},
  numpages = {4},
  year = {2003},
  month = {Jul},
  publisher = {American Physical Society},
  doi = {10.1103/PhysRevLett.91.017001},
  url = {https://link.aps.org/doi/10.1103/PhysRevLett.91.017001}
}

@ARTICLE{Pol69rmp,
  author = {Pollack, G. L.},
  title = {Kapitza Resistance},
  journal = {Rev. Mod. Phys.},
  year = {1969},
  volume = {41},
  pages = {48--81},
  month = {Jan},
  doi = {10.1103/RevModPhys.41.48},
  issue = {1},
  numpages = {0},
  publisher = {American Physical Society},
  url = {https://link.aps.org/doi/10.1103/RevModPhys.41.48}
}

@ARTICLE{Sid18prb,
  author = {Sidorova, Mariia V. and Kozorezov, A. G. and Semenov, A. V. and Korneeva,
	Yu. P. and Mikhailov, M. Yu. and Devizenko, A. Yu. and Korneev, A.
	A. and Chulkova, G. M. and Goltsman, G. N.},
  title = {Nonbolometric bottleneck in electron-phonon relaxation in ultrathin
	{WSi} films},
  journal = {Phys. Rev. B},
  year = {2018},
  volume = {97},
  pages = {184512},
  month = {May},
  doi = {10.1103/PhysRevB.97.184512},
  issue = {18},
  numpages = {13},
  publisher = {American Physical Society},
  url = {https://link.aps.org/doi/10.1103/PhysRevB.97.184512}
}

@ARTICLE{Zol14ltp,
  author = {Zolochevskii, I. V.},
  title = {Resistive states in wide superconducting films induced by dc and
	ac currents (Review Article)},
  journal = {Low Temp. Phys.},
  year = {2014},
  volume = {40},
  pages = {867-892},
  number = {10},
  doi = {http://dx.doi.org/10.1063/1.4900695},
  url = {http://scitation.aip.org/content/aip/journal/ltp/40/10/10.1063/1.4900695}
}

@ARTICLE{Lar76etp2,
  author = {Larkin, A. and Ovchinnikov, Yu.},
  title = {Nonlinear Conductivity of Superconductors in the Mixed State},
  journal = {Sov. Phys. JETP},
  year = {1976},
  volume = {41},
  pages = {960}
}

@article{Bez19prb,
  author       = {Alexei Bezuglyj and Valerij Shklovskij and Ruslan Vovk and Volodymyr Bevz and Michael Huth and Oleksandr Dobrovolskiy},
  title        = {Local flux-flow instability in superconducting films near {$T_c$}},
  journal      = {Phys. Rev. B},
  volume       = {99},
  number       = {17},
  pages        = {174518},
  year         = {2019},
  doi          = {10.1103/PhysRevB.99.174518}
}

@BOOK{Tin04boo,
  title = {Introduction to Superconductivity},
  publisher = {Mineola, New York},
  year = {2004},
  author = {Tinkham, M.}
}

@article{Rui26pms,
  author    = {Ruiz \emph{et al.}, H.},
  title     = {Critical current density in advanced superconductors},
  journal   = {Progr. Mater. Sci.},
  volume    = {155},
  pages     = {101492},
  year      = {2026},
  doi       = {10.1016/j.pmatsci.2025.101492},
  url       = {https://www.sciencedirect.com/science/article/pii/S0079642525000702}
}

@Article{Emb17nac,
author={Embon \emph{et al.}, L.},
title={Imaging of super-fast dynamics and flow instabilities of superconducting vortices},
journal={Nat. Commun.},
year={2017},
day={20},
volume={8},
number={1},
pages={85},
issn={2041-1723},
doi={10.1038/s41467-017-00089-3},
url={https://doi.org/10.1038/s41467-017-00089-3}
}

@article{Pfa24apl,
author = {Pfaff, C. and Petit-Watelot, S. and Andrieu, S. and Pasquier, L. and Ghanbaja, J. and Mangin, S. and Dumesnil, K. and Hauet, Thomas},
title = {Spin injection at {MgB}$_{2}$-superconductor/ferromagnet interface},
journal = {Appl. Phys. Lett.},
volume = {125},
number = {10},
pages = {102601},
year = {2024},
month = {09},
issn = {0003-6951},
doi = {10.1063/5.0220815}
}

@article{Bra95rpp,
doi = {10.1088/0034-4885/58/11/003},
url = {https://doi.org/10.1088/0034-4885/58/11/003},
year = {1995},
month = {nov},
publisher = {},
volume = {58},
number = {11},
pages = {1465},
author = {Brandt, E. H.},
title = {The flux-line lattice in superconductors},
journal = {Rep. Prog. Phys.},
}

@article{Dob19pra,
  title = {Fast Dynamics of Guided Magnetic Flux Quanta},
  author = {Dobrovolskiy, O.V. and Bevz, V.M. and Begun, E. and Sachser, R. and Vovk, R.V. and Huth, M.},
  journal = {Phys. Rev. Appl.},
  volume = {11},
  issue = {5},
  pages = {054064},
  numpages = {9},
  year = {2019},
  month = {May},
  publisher = {American Physical Society},
  doi = {10.1103/PhysRevApplied.11.054064},
  url = {https://link.aps.org/doi/10.1103/PhysRevApplied.11.054064},
}

@Article{Kunchur2002,
  author     = {Kunchur, Milind N.},
  journal    = {Physical Review Letters},
  title      = {Unstable Flux Flow due to Heated Electrons in Superconducting Films},
  year       = {2002},
  issn       = {1079-7114},
  month      = sep,
  number     = {13},
  pages      = {137005},
  volume     = {89},
  comment    = {Kunchur model of Flux-Flow Instability, FFI at low temperatures arises from expansion of the vortices and increase in quasiparticle population, as opposed to Larkin-Ovchinnikov (LO) theory. At lowe},
  doi        = {10.1103/physrevlett.89.137005},
  publisher  = {American Physical Society (APS)},
  ranking    = {rank5},
  readstatus = {skimmed},
}

@Article{Kas18aip,
author={Kasaei, L.
and Melbourne, T.
and Manichev, V.
and Feldman, L. C.
and Gustafsson, T.
and Chen, Ke
and Xi, X. X.
and Davidson, B. A.},
title={{MgB2} {Josephson} junctions produced by focused helium ion beam irradiation},
journal={AIP Adv.},
year={2018},
month={Jul},
day={19},
volume={8},
number={7},
pages={075020},
doi={10.1063/1.5030751}
}

\end{document}